\documentclass[aps,preprint,groupedaddress,floatfix]{revtex4}
\usepackage{graphicx}
\begin{document}

\title{ Electro-weak production of pseudovector C-even heavy quarkonia in electron-positron collisions on Belle II and BES III}
\author {
Nikolay Achasov }

\affiliation{ Laboratory of Theoretical Physics,
 Sobolev Institute for Mathematics, 630090, Novosibirsk, Russia
}


\begin{abstract}
\begin{enumerate}
\item The $X(3872)$ State as  Charmonium $\chi_{c1}(2P)$
\item  The  $\chi_{c1}$ and  $\chi_{b1}$  production  in the
$e^+e^-\to\chi_{c1}/\chi_{b1}$ reaction
\end{enumerate}
\end{abstract}

\maketitle

\section{OUTLINE}
\label{outline} 
\begin{figure}
\begin{center}
\hspace*{-3cm}\includegraphics[height=16cm,width=8cm,angle=270]{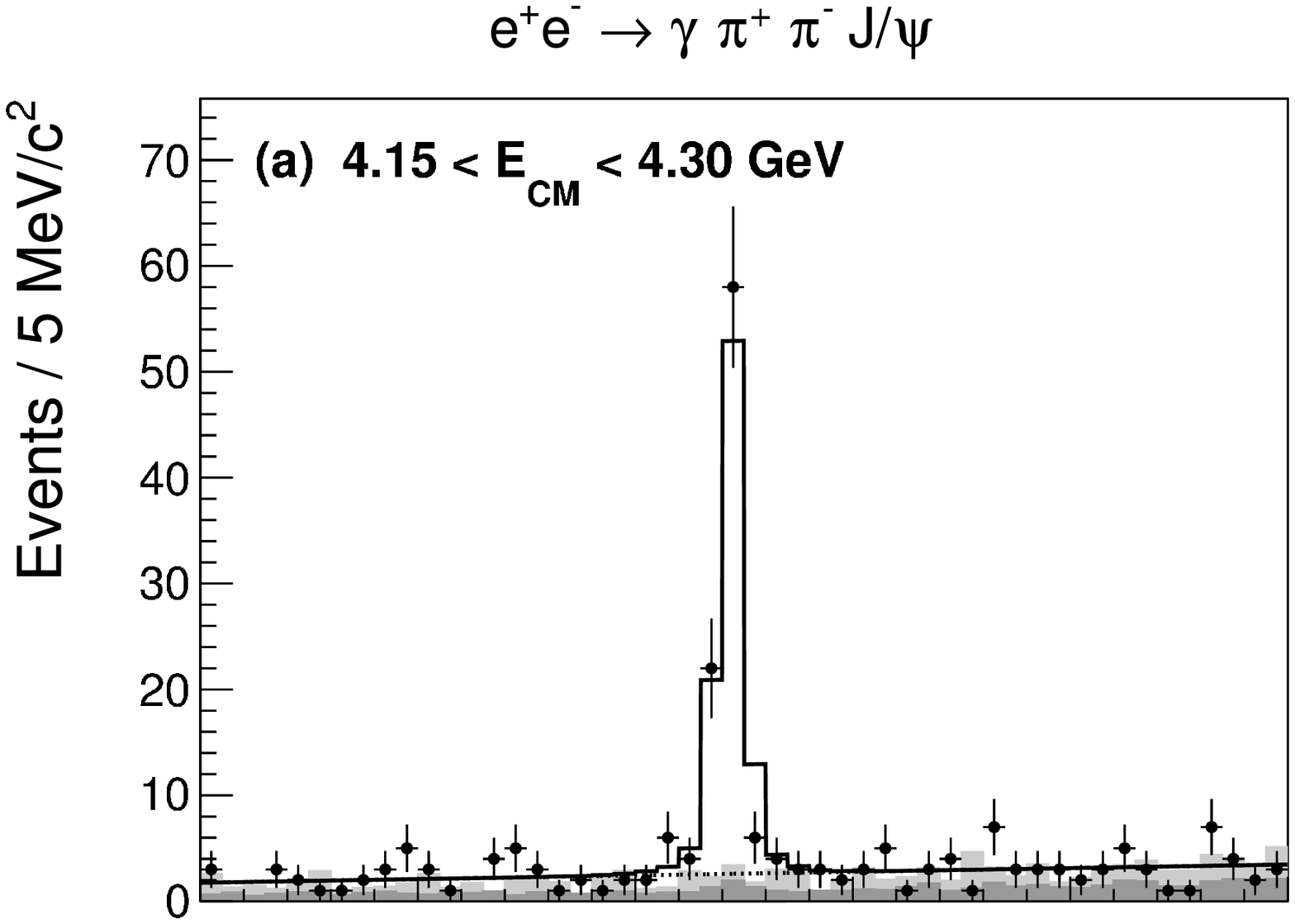}
\end{center}
\caption{ Distribution of $\pi^+\pi^- J/\psi$ mass from Ref.
\cite{BESIII2019}.
 }
\label{Data1}
\end{figure}

\begin{figure}
\begin{center}
\hspace*{-3cm}\includegraphics[height=16cm,width=8cm,angle=270]{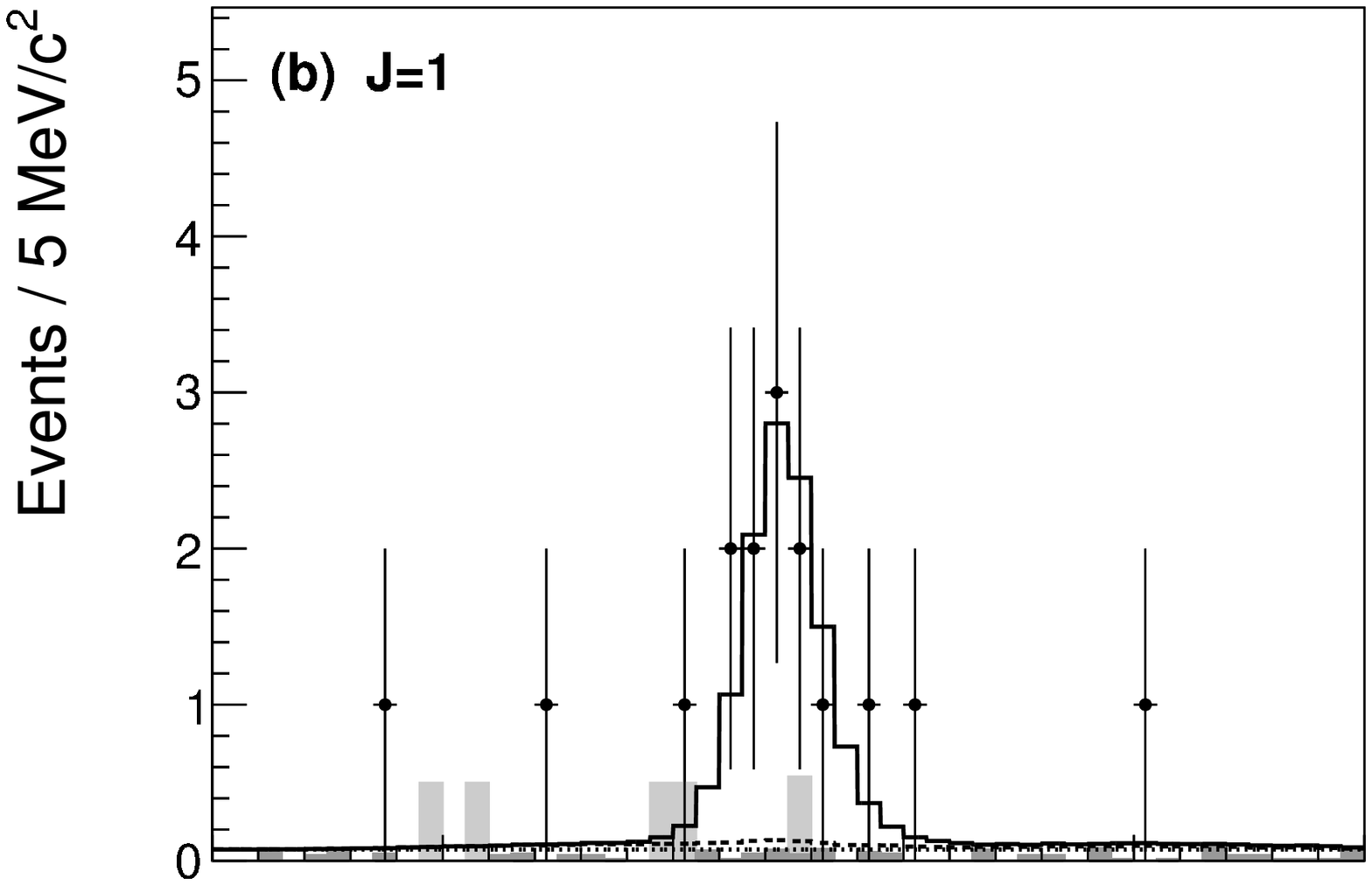}
\end{center}
\caption{ Distribution of $\pi^0\chi_{c1}(1P)$ mass from
\cite{BESIII2019}.
 }
\label{Data2}
\end{figure}

The $X(3872)=\chi_{c1}(3872)$ meson
\cite{pdg-2018}, a patriarch of the $XYZ$ spectroscopy,  was
appointed to be the $D^0\bar D^{*0} + c.c.$ molecule  with a
radius greater than 3 fermi from the very beginning despite the
fact that $X(3872)=\chi_{c1}(3872)$ is produced  in hard processes
with a radius less than one fermi as intensively as the compact
charmonium $\psi(2S)$. Even  the landmark result of the LHCb
Collaboration  \cite{LHCb14}
\begin{equation}
\label{Xtogamma}
 \frac{BR(X\to\gamma \psi(2S))}{BR(X\to\gamma
J/\psi)}=2.46\pm 0.7\,,
\end{equation}
directly pointing to the charmonium nature of $X(3872)$,  did not
stop the molecular lobby.
\\[6pt]  We reviewed the scenario in
detail where $X(3872)$ resonance is the $c\bar c=\chi_{c1}(2P)$
charmonium which "sits on" the $D^0\bar D^{*0}$ threshold. We
explained all known data on $X(3872)$ and suggested clear program
of verification of  our scenario \cite{NNA+EVR,NNA+EVR+,NNA3872}.
\\[6pt]
 We predicted the
significant number of decay channels via two gluons: $X(3872)\to
gluon\, gluon\to light\, hadrons$,  the same as in the case
$\chi_{c1}(1P)\to gluon\, gluon\to light\, hadrons $.
 It means that two virtual gluons can produced  the $X(3872)$
 resonance
 \begin{equation}
 \label{2}
 e^+e^-\to\psi(m_i)\to\gamma\,gluon\,gluon\to\gamma X(3872),
 \end{equation}
 here $\psi(m_k)$: $I^G(J^{PC})=0^-(1^{--})\,,\ m_i>m_{X(3872)}$.
The BES III Collaboration  found the $X(3872)$ resonance in the
reaction $e^+e^-\to\gamma X(3872)$ at center-of-mass energies for
4.009 to 4.420  GeV \cite{BESIII2014}
\begin{eqnarray}
\label{3}
 && e^+e^-\to\sum_i\psi(m_i)\to\gamma\,gluon\,gluon\to\gamma
X(3872)=\nonumber\\ &&  =  \psi(4040)+
\psi(4160)+\psi(4230)+\psi(4260)+\psi(4360)+ \nonumber\\
 && +
\psi(4390)+\psi(4415)
 \to\gamma\,gluon\,gluon\to\gamma\, X(3872)\to\nonumber \\
&& \to\gamma\pi^+\pi^- J/\psi\,.
\end{eqnarray}
  Recently the BES III Collaboration found  the $X(3872)$ resonance
  in the reaction $e^+e^-\to\gamma X(3872)$ at  center-of-mass energies for 4.15 to
4.3  GeV \cite{BESIII2019}
\begin{eqnarray}
\label{4} &&
e^+e^-\to\psi(4160)+\psi(4230)+\psi(4260)\to\nonumber\\
 && \to\gamma\,gluon\,gluon\to \gamma\,
 X(3872)\to\gamma\pi^+\pi^- J/\psi\,,
\end{eqnarray}
see Fig. \ref{Data1}  and
\begin{eqnarray}
\label{5} &&
e^+e^-\to\psi(4160)+\psi(4230)+\psi(4260)\to\nonumber\\
 && \to\gamma\,gluon\,gluon\to \gamma\,
 X(3872)\to\gamma\pi^0\chi_{c1}(1P)\,,
\end{eqnarray}
see Fig. \ref{Data2}.\\[6pt]

 The giant colourless molecule does not connected
with gluons! Its colourless  constituents $D^0\,,\bar D^{*0}$ do
not connected with gluons also! So The BES III Collaboration
closes the molecular model of the $X(3872)$ resonance.\\[6pt]
 As for the tetraquark model, the two-gluon
production of the $X(3872)$ resonance is possible
$e^+e^-\to\gamma\,gluon\,gluon\to\gamma q\bar q c\bar c\to\gamma
X(3872), q=u, d$. But, such a process is described by nonplanar
diagrams, which are depressed always.  So the BES III
collaboration  puts in a difficult position the tetraquark model
of the $X(3872)$ resonance.\\[6pt] Thus  the BES III collaboration
confirms the $c\bar c$ charmonium model
  of the  $X(3872)$ resonance\,.
 \begin{center}
{\bf\boldmath{$ $} $X(3872)=\chi_{c1}(2P)$\,! }
 \end{center}
 It is often thought that violations of isotopic
invariance in the decays\\ $X(3872)\to\pi^+\pi^- J/\psi$ and
$X(3872)\to\pi^0\chi_{c1}(1P)$ are crucial for the $X(3872)$
nature.\\ However, this is a misunderstanding. These are the
problems of the second row.\\[6pt] The point is that
electromagnetic interaction is not small in this energy region,
$\alpha\sim\alpha_s^3$. As a result $BR(J/psi\to g\,g\,g)=
(64.1\pm 1.0)\,\%\,, \ BR(J/psi\to\gamma\,\,g\,g)= (8.8\pm
1.1)\,\%\,,\   BR(J/psi\to\,\ virtual\,\ \gamma\to hadrons)=
(13.5\pm 0.30)\,\%$ \cite{pdg-2018}\,.  Close to our scenario is
an example of the $J/\psi\to\rho\eta'$ and $J/\psi\to\omega\eta'$
decays. According to Ref. \cite{pdg-2018}
\begin{equation}
\label{Jpsi}
 BR(J/\psi\to\rho\eta')=(1.05\pm 0.18)\cdot10^{-4}\ \ \mbox{and}\ \
BR(J/\psi\to\omega\eta')=(1.82\pm 0.21)\cdot10^{-4}.
\end{equation}
The similar picture is shown also by the $\psi(2S)$
\cite{pdg-2018}\,.
\\[6pt]
As for the isotopic symmetry violation  via $m_d-m_u$, it can be
considerable also, for example,  the $\rho^0 -\omega$ and
$\eta-\pi^0$ transitions\,  are of the order $(m_d-m_u)\times 1$
GeV order \cite{pi0eta}.\\[6pt]
 As for $X(3872)\to\pi^+\pi^- J/\psi$ and $X(3872)\to\omega J/\psi$,
this problem is discussed in detail in Refs.
\cite{NNA+EVR+,NNA3872}. It is interesting to note that else in
Ref. \cite{2005} there was shown that the $\omega$ do'not produced
virtually in the $X(3872)\to\omega J/\psi$ decay. One can see only
the left tail of $\omega$ far off the resonance. Let us add that a
background can interfere with this tail constructively or
destructively. But the molecular lobby hard discusses the strong
isotopic breaking in the above decays.\\[6pt] As for
$X(3872)\to\pi^0\chi_{c1}(1P)$, it is possible a such scheme\\
$X(3872)\to gluon\,gluon\,\chi_{c1}(1P)\to\eta\chi_{c1}(1P)\to
\pi^0\chi_{c1}(1P)$ via $\eta -\pi^0$ mixing.\\[6pt]
 I dare
recommend looking for the decays $\chi_{b1}(2P)\to\rho^0 J/\psi$
and $\chi_{b1}(2P)\to\pi^0\chi_{b1}(1P)$\,.
\section{OUTLOOK}
\label{outlook}

\begin{figure}
\begin{center}
\includegraphics[height=4cm,width=10cm]{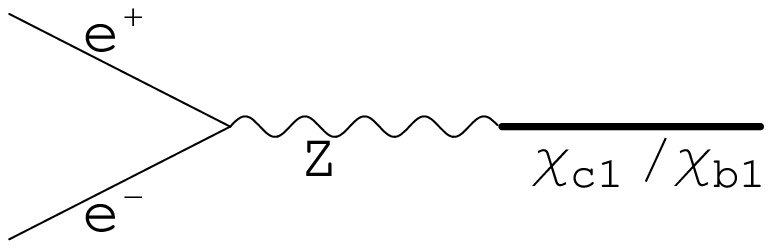}
\end{center}
\caption{ } \label{Theory1}
\end{figure}

 In this energy region the weak
interaction grows with energy increase $\propto G_FE^2$, here
$G_F=10^{-5}m_p^{-2}$ is the Fermi constant.\\[6pt]
$G_FE^2=1.4\times 10^{-4}$ for $\chi_{c1}(1P)$ and
$G_FE^2=1.7\times 10^{-4}$ for $\chi_{c1}(3872)$. That is,
$G_FE^2\sim \alpha^2$ in the BES III energy region.\\[6pt]
 $G_FE^2=1.1\times 10^{-3}$ for $\chi_{b1}(1P)$ and
$G_FE^2=1.2\times 10^{-3}$ for $\chi_{b1}(2P)$. That is,
$G_FE^2\gg\alpha^2$ in the Belle II energy region.\\[6pt]
 The
BESS III luminosity $10^{33}cm^{-2}s^{-1}$ gives possibilities to
register near hundred of events of the $e^+e^-\to
Z\to\chi_{c1}(1P)$ decays, Fig.\,\ref{Theory1},  per day
\cite{1996} and near thirty of them in the well-known channel
$\chi_{c1}(1P)\to\gamma\psi(1S)$. If
$\chi_{c1}(3872)=\chi_{c1}(2P)$,
 then also near hundred of events of the
 $e^+e^-\to Z\to\chi_{c1}(3872)$ decays, Fig.\,\ref{Theory1},
 per day may be registered and
several of them in the channel $\chi_{c1}(3872)\to\gamma\psi(2S)$,
several tens of them in the channel
 $\chi_{c1}(3872)\to D^0\bar
D^{*0}+c.c.$.\\[6pt ] The huge Belle II luminosity $8\times
10^{35}cm^{-2}s^{-1}$ gives possibilities to register near hundred
thousand of events the each $e^+e^-\to Z\to\chi_{b1}(1P)$ and
$e^+e^-\to Z\to\chi_{b1}(2P)$ decays, Fig.\,\ref{Theory1}, per day
\cite{1996} and several tens of thousands of them in the
well-known channels $\chi_{b1}(1P)\to\gamma\Upsilon(1S)$ and
$\chi_{b1}(2P)\to\gamma\Upsilon(2S)$.\\[6pt]
 Note that the above
estimations were became under the assumption that the resonance
widths are not small compared to the energy resolution. \\[6pt]

\begin{figure}
\begin{center}
\includegraphics[height=4cm,width=10cm]{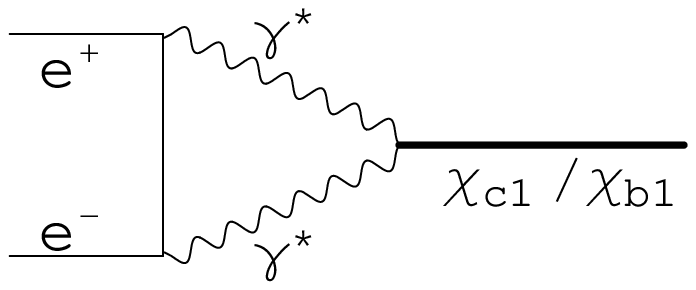}
\end{center}
\caption{ } \label{Theory2}
\end{figure}

In the Belle II energy region  will dominate  the one-Z-boson
mechanism, Fig. \ref{Theory1}. As for BESS III energy region, the
one-Z-boson mechanism, Fig. \ref{Theory1}, and the two-photon
mechanism, Fig. \ref{Theory2}, are probably of the same order.
Fortunately, the $e^+e^-\to Z\to\chi_{c1}/\chi_{b1}$ and
$e^+e^-\to\gamma^{\ast} \gamma^{\ast}\to\chi_{c1}/\chi_{b1}$
contributions do not interfere in the total cross sections. Note
that the $\gamma^{\ast}\gamma^{\ast}\chi_{c1}/\chi_{b1}$ vertex
was considered  in Refs. \cite{1992-1993,1994}\,in details. The
creation of longitudinally polarized electron-positron beams
allows to study both the total cross sections and the interference
of $e^+e^-\to Z\to\chi_{c1}/\chi_{b1}$ with
$e^+e^-\to\gamma^{\ast}\gamma^{\ast}\to\chi_{c1}/\chi_{b1}$. More
specific
\begin{eqnarray}
\label{amplitudes} && e^+(1)e^-(-1)\to\chi_{c1}/\chi_{b1}(-1) =
e^+(1)e^-(-1)\to
 Z\to\chi_{c1}/\chi_{b1}(-1) +\nonumber\\
 && + e^+(1)e^-(-1)\to\gamma^{\ast}\gamma^{\ast}\to\chi_{c1}/\chi_{b1}(-1)\,,
 \nonumber\\[6pt]
 && e^+(-1)e^-(1)\to\chi_{c1}/\chi_{b1}(1) =
-\ \ (!)\ \ e^+(-1)e^-(+1)\to
 Z\to\chi_{c1}/\chi_{b1}(1) +\nonumber\\
 && +
 e^+(-1)e^-(1)\to\gamma^{\ast}\gamma^{\ast}\to\chi_{c1}/\chi_{b1}(1)\,,
\end{eqnarray}
where $\lambda$ in $e^+(\lambda)$\,, $e^-(\lambda)$\,, and
$\chi_{c1}/\chi_{b1}(\lambda)$ is the projection of the particle
spin on the electron momentum direction in the mass center system.
We neglected $m_e$.

\section{SUMMARY}
The new elegant experimental probes appear. In particular, they
could find out whether is $\chi_{c1}(3872)=\chi_{c1}(2P)$ and
search out the $\chi_{b1}(2P)\to\rho^0\Upsilon(1S)$ and
$\chi_{b1}(2P)\to\pi^0\chi_{b1}(1P)$.

\section{Acknowledgments}

   I am grateful to
Organizers of {\bf From Phi to Psi - 2019} for the kind
Invitation.

The work was supported by the program No. II.15.1 of fundamental
scientific researches of the Siberian Branch of Russian Academy of
Sciences, the project No. 0314-2019-0021.


\begin{thebibliography}{99}
\bibitem{pdg-2018}
M. Tanabashi et al. (Particle Data Group), Phys. Rev. D {\bf 98},
030001 (2018).

\bibitem{LHCb14} R. Aaij, et al. ( LHCb Collaboration), Nucl.
Phys. B {\bf 886}, 665 (2014).
\bibitem{NNA+EVR}
 N.N. Achasov and E.V. Rogozina, JETP Lett. {\bf 100}, 227 (2014).

\bibitem{NNA+EVR+}
N.N. Achasov and E.V. Rogozina, Mod. Phys. Lett. A {\bf 30},
1550181 (2015);\\
 J.Univ.Sci.Tech.China {\bf 46}, 574 (2016).

\bibitem{NNA3872}
 Nikolay Achasov, EPJ Web Conf. {\bf 125}, 04002 (2016);\\
N.N. Achasov, Phys. Part. Nucl. {\bf 48}, 839 (2017);\\
 Nikolay Achasov, EPJ Web Conf. {\bf 191}, 04002 (2018) .

\bibitem{BESIII2014}  M. Ablikim {\it et al.} (BESIII Collaboration), Phys. Rev. Lett. {\bf 112},
092001(2014).

\bibitem{BESIII2019}  M. Ablikim {\it et al.} (BESIII Collaboration), arXiv: 1901.03992 v1 [hep-ex] 13 Jan 2019.

\bibitem{pi0eta} B.L. Ioffe,  Phys. Usp. {\bf 44} 1211 (2001); Usp. Fiz. Nauk {\bf 171} 1273
(2001).

\bibitem{2005} Mahiko Suzuki, Phys. Rev. D {\bf 72}, 114013 (2005).

\bibitem{1996} N.N. Achasov, JETP Lett. {\bf 63}, 601 (1996);\\
arXiv:hep-ph/9603262v1; arXiv:hep-ph/9604226v2.

\bibitem{1992-1993} N. N. Achasov, Phys. Lett. B {bf 287}, 213 (1992);\\
Sov.Phys. JETP {\bf 74}, 913 (1992)\,[ZhETF {\bf 101}, 1713
(1992)];\\ JETP Lett. {\bf 56}, 329 (1992);\\ JETP {\bf76}, 5
(1993)\,[ZhETF {\bf 103}, 11 (1993)].

\bibitem{1994} N. N. Achasov, Particle World, {\bf Vol. 4}, No. 10, pp 10-13 (1994).


\end{thebibliography}
\end{document}